\documentstyle[12pt]{article}
\newcommand{\hc}{h_{v}^{(c)}}                
\newcommand{\hcbar}{\bar h_{v'}^{(c)}}        
\newcommand{\hcbarv}{\bar h_{v}^{(c)}}        
\newcommand{\vslash}{\mbox{$\not{\hspace{-1.03mm}v}$}}        
\newcommand{\Dslash}{\mbox{$\not{\hspace{-1.03mm}D}$}}        
\newcommand{\epsslash}{\mbox{$\not{\hspace{-1.03mm}\epsilon}$}} 
\newcommand{\bea}{\begin{eqnarray}}
\newcommand{\eea}{\end{eqnarray}}
\newcommand{\beq}{\begin{equation}}
\newcommand{\eeq}{\end{equation}}
\newcommand{\bay}{\begin{array}}
\newcommand{\eay}{\end{array}}
\begin{document}
\begin{titlepage}
\begin{flushleft}
MZ-TH/92-51\\
November 1992
\end{flushleft}
\begin{center}
\large
\bf
{\Large Radiative Decays of the P-wave Charmed Mesons}\\[2cm]
\rm
J.G.K\"orner$^1$, D.Pirjol$^2$, K.Schilcher$^1$\\[.5cm]
Institut f\"ur Physik, Staudinger Weg 7\\
Johannes Gutenberg-Universit\"at\\
Postfach 3980, D-6500 Mainz\\
Germany\\[2cm]
\normalsize
\bf
Abstract\\
\rm
\small
\end{center}
   The predictions of the large mass limit for the radiative decays of
the known p--wave charmed mesons are analyzed.Special attention is
devoted to the problem of gauge invariance of the transition matrix
elements.The width ratios arising from heavy quark symmetry are
given for the different multipole components.Finally, some estimates
for the rates of these decays are given, using the constituent
quark model of Isgur, Scora, Grinstein and Wise.\\
[3cm]
\footnotesize
$^1\,$Supported in part by the BMFT, FRG under contract 06MZ730\\
$^2\,$Supported by Graduiertenkolleg Teilchenphysik, Universit\"at Mainz\\
\normalsize
\end{titlepage}

\section*{1.Introduction}

   The strong, weak and electromagnetic interactions of mesons containing
one heavy quark can be described in terms of an effective theory
which is simultaneously invariant under heavy quark spin--flavour
symmetry, the chiral $SU_L(2) \times SU_R(2)$ and the $U_{e.m.}(1)$
gauge group. This effective theory has recently been applied \cite{1,2}
to predict the $B^* \to B\gamma$ decay widths from the CLEO data\cite{3}
on branching ratios for $D^{*+} \to D\gamma (\pi )$. There is, as yet,
no similar data on radiative decays of the known p--wave D mesons,
although the mode $D_{s_{1}}^*(2536) \to D_s^{*+}\gamma$ is reported
\cite{4} as having possibly been seen. However, the situation might change
soon if enough statistics becomes available. It is therefore of some
interest to see what information can be obtained about these decays by
using solely the symmetries of the large mass limit. In the present paper
we study the radiative decays of the members of the $s_{\ell}^{\pi_{\ell}}
 = \frac{3}{2}^+$ multiplet of charmed mesons to the respective charmed
s-wave states. Previous work on the same problem has been done in
\cite{SIMILAR1,SIMILAR2,SIMILAR3}. These authors
employed a specific dual model and tensor meson dominance and/or made use
of the $SU(4)$ flavour symmetry to extract the necessary couplings. Such an
approach is clearly not compatible with the large mass effective theories
referred to above. We prefer instead to describe the electromagnetic
interactions of these mesons using a method which explicitly displays
the new spin and flavour symmetries appearing in the heavy mass limit.
We show that the
heavy quark contribution to the matrix elements of these transitions
is, up to ${\cal O}(1/m_c)$, completely determined in terms of the
same Isgur--Wise function $\xi_{3/2}^*(v\cdot v')$ which describes
the semileptonic decays of the $\bar B$ mesons to excited $s_{\ell}
^{\pi_{\ell}} = \frac{3}{2}^+$ p--wave
D--mesons\cite{5}\footnote{More precisely, this is only true for the
$E1$ component of the transition.}. Ratios of partial widths
are given for the different multipole components. Finally, the constituent
quark model of Isgur, Scora, Grinstein and Wise \cite{6} is
used in order to make some estimates for the widths of these decays.
The result is that the  decays of the $D_s^+$ mesons are dominated by the
magnetic quadrupole $M2$ mode while the $D^0$ mesons decay predominantly
through an electric dipole $E1$ mode.

\section*{2.The method}

   We are interested in the matrix elements of the electromagnetic
current
\beq
 J_{e.m.}^{\mu} = \frac{2}{3}\bar u\gamma^{\mu} u
                - \frac{1}{3}\bar d\gamma^{\mu} d
                - \frac{1}{3}\bar s\gamma^{\mu} s
                + \frac{2}{3}\bar c\gamma^{\mu} c
\eeq
taken between one of the s--wave charm  meson states and one of the two
degenerate states of the $s_{\ell}^{\pi_{\ell}} = \frac{3}{2}^+$ p--wave
multiplet:
\beq
 {\cal M}_{ij}^{\mu} = \langle M_i(v')|J_{e.m.}^{\mu}|M^{**}_j(v)\rangle,
\eeq
where $M_i(v'), M^{**}_j(v)$ are generic notations for the respective
meson states,\\
 $(M_1 , M_2)=(D , D^*)$ and $(M_1^{**} , M_2^{**})=
(D_1 , D_2^*)$. Their fields can be combined, as usual \cite{USUAL} into
a Dirac
matrix\footnote{A factor of $\sqrt{M_S} (\sqrt{M_{P_{3/2}}})$ is
absorbed into the field $H$ ($T^{\alpha}$).}
\bea
 H &=& \frac{1+\vslash\, '}{2}\lbrack D^{*\lambda}\gamma_{\lambda} -
                                  D\gamma_5\rbrack  \\
 T^{\alpha} &=& \frac{1+\vslash }{2}\lbrace D_2^{*\alpha\lambda}
  \gamma_{\lambda} - \frac{1}{\sqrt{6}}D_1^{\lambda}\gamma_5
  \lbrack 3g^{\alpha}_{\lambda} - \gamma_{\lambda}(\gamma^{\alpha}-
  v^{\alpha})\rbrack\rbrace .
\eea

   We will choose to evaluate the matrix element (2) of the first three
terms in (1) (the light quarks' electromagnetic current) in a somewhat
different way from the last one (the heavy quark electromagnetic current).
Namely, the former is given in the effective theory by the matrix elements
of the most general operator which includes the fields (3-4), has positive
parity and is gauge invariant. To second order in the photon momentum
it can be written as
\bea
\lefteqn{ n_{\mu}\tilde J^{\mu}_{light} =
 f_1\mbox{Tr}\lbrack \bar H T^{\alpha}\gamma^{\beta}F_{\alpha\beta}\rbrack
+f_2\mbox{Tr}\lbrack \bar H T^{\alpha} v^{\beta}F_{\alpha\beta}\rbrack}
\\ \nonumber
&&+\frac{g_1}{\Lambda_{\chi}}\mbox{Tr}\lbrack\bar H T^{\alpha}
\partial_{\lambda}\sigma^{\lambda\beta}F_{\alpha\beta}\rbrack
+\frac{g_2}{\Lambda_{\chi}}\mbox{Tr}\lbrack \bar H T^{\alpha}
\partial_{\alpha}\sigma_{\mu\nu} F^{\mu\nu}\rbrack
+\frac{ig_3}{\Lambda_{\chi}}\mbox{Tr}\lbrack \bar H T^{\alpha}
(\gamma^{\lambda}v^{\beta}+\gamma^{\beta}v^{\lambda})\partial_{\lambda}
F^{\mu\nu}\rbrack\,.
\eea
   Here $F^{\mu\nu}=\partial^{\mu}A^{\nu}-\partial^{\nu}A^{\mu}$ is the
electromagnetic field tensor and $n_{\mu}$ is the polarization vector of
the emitted photon. Note that the velocities of the two heavy meson fields
in (5) must be taken to be equal $v=v'$, since otherwise the invariance
under the $SU_{spin}(2)$ heavy quark spin symmetry group is lost. Terms
with one derivative acting on the heavy meson fields are forbidden by
reparametrization invariance constraints, as shown in \cite{7} for the
case of the strong interactions of the same mesons. $f,g,h$ and $j$ are
unknown constants which have to be determined from a comparison with
experiment. $\Lambda_{\chi}$=1 $GeV$ suppresses terms with one derivative
on the photon field.

   The matrix element (2) of the last term in (1) will be evaluated
by using usual HQET methods \cite{8}. The vector current $\bar c
\gamma_{\mu} c$ is matched in the HQET to ${\cal O}(1/m_c)$ onto the
operator
\bea
 \tilde J_{heavy}^{\mu} = e^{im_c (v'-v)\cdot x}
\lbrace \hcbar\gamma^{\mu}\hc - \frac{i}{2m_c}\hcbar\lbrack
\stackrel{\leftarrow}{\Dslash}\gamma^{\mu}-\gamma^{\mu}
\stackrel{\rightarrow}{\Dslash}\rbrack\hc\rbrace\,.
\eea
   The two velocities $v, v'$ are taken to be different here and satisfy
\beq
  M_{P_{3/2}} v_{\mu} = M_S v'_{\mu} + k_{\mu}
\eeq
with $k_{\mu}$ the photon momentum. Defining the binding energies
\beq
 \bar\Lambda' = M_S-m_c\, , \qquad \bar\Lambda = M_{P_{3/2}} - m_c\, ,
\eeq
we have
\bea
v\cdot k &=& (\bar\Lambda-\bar\Lambda')\lbrack 1-\frac{\bar\Lambda-
\bar\Lambda'}{2m_c}+\frac{\bar\Lambda(\bar\Lambda-\bar\Lambda')}{2m_c^2}
+{\cal O}(\frac{\bar\Lambda^3}{m_c^3},\frac{\bar\Lambda'^3}{m_c^3})
\rbrack \\
v\cdot v' &=& 1+\frac{(\bar\Lambda-\bar\Lambda')^2}{2m_c^2}
+{\cal O}(\frac{\bar\Lambda^3}{m_c^3},\frac{\bar\Lambda'^3}{m_c^3}).
\eea
   It can be seen from (10) that $v\cdot v'$ differs from 1 only through
terms of ${\cal O}(1/m_c^2)$. This provides the justification for our
assumption about the equality of the two velocities $v$ and $v'$ built
into the equation (5). However, in considering the matrix elements of the
heavy quark current (6), we will stick to a two--velocities description,
as a means for introducing the photon momentum $k_{\mu}$ {\em via} (7).

   Using the usual trace formalism, we have
\bea
\lefteqn{\langle M_i(v')|\tilde J_{heavy}^{\mu} (x)|M_j^{**}(v)\rangle=}
\\ \nonumber
&&e^{i(M_S v'-M_{P_{3/2}}v)\cdot x}\sqrt{M_S M_{P_{3/2}}} \mbox{Tr}
\lbrack \hat M_i^{\dagger}\frac{1+\vslash\, '}{2}\gamma^{\mu}
\frac{1+\vslash}{2}\hat M_j^{**\alpha}v'_{\alpha}\rbrack
\xi_{3/2}^*(v\cdot v'),
\eea
where only the leading term in (6) has been retained. $\hat M_i$ and
$\hat M_j^{**\alpha}$ are the usual interpolating fields associated
with the respective meson fields:
\bea
 (\hat M_1^{\dagger}\,\, ,\,\, \hat M_2^{\dagger}) &=& (\gamma_5\,\,
 , \,\,\epsslash\,^*), \\
 (\hat M_1^{**\alpha}\,\, , \,\,\hat M_2^{**\alpha}) &=& \left( -\frac{1}
{\sqrt{6}}\epsilon^{\lambda}\gamma_5\lbrack 3g_{\lambda}^{\alpha} -
\gamma_{\lambda}(\gamma^{\alpha}-v^{\alpha})\rbrack \,\, , \,\,
\epsilon^{\alpha\lambda}\gamma_{\lambda} \right) .
\eea
   Unfortunately, the matrix element (11) is not gauge invariant.
Contracting equ. (11) with $k_{\mu}=M_{P_{3/2}}v_{\mu} - M_Sv_{\mu}'$
gives a result of order ${\cal O}(m_c^0)$, proportional to $(\bar\Lambda-
\bar\Lambda')$. It is not difficult to trace the origin of this difficulty:
it is the nonconservation of the HQET vector current (6) when only
the first term is retained. In order to have vector current conservation,
the second term in (6) must be included as well. Only then we have
\beq
       \partial_{\mu}\tilde J^{\mu}_{heavy} = {\cal O}(1/m_c) \,.
\eeq

   Of course, vector current conservation holds true up to any given
order in $1/m_c$, provided the necessary additional terms are added to
(6).

   On the other hand, the inclusion of the second term in (6) requires
also consideration of the $\frac{1}{m_c}$--terms in
the heavy quark effective lagrangian
\bea
  {\cal L}_{HQET} = \hcbarv (iv\cdot D)\hc
                + \frac{1}{2m_c}\hcbarv (iD)^2\hc
          - \frac{g}{4m_c}\hcbarv\sigma\cdot G\hc \,,
\eea
which could give in principle contributions of comparable magnitude.
As is well known since Luke's paper\cite{LUKE} all these bring along
new, unknown form--factors. We will show, however, that to first order
in $\frac{v\cdot k}{m_c}$, {\em no} new form--factors show up besides
$\xi_{3/2}^*(v\cdot v')$.

   We shall first evaluate the contribution of the second term in equ.(6).
To this end consider the matrix elements of the operator $-i\hcbar
\stackrel{\leftarrow}{D}_{\rho}\Gamma\hc$, which are given by
\bea
\lefteqn{\langle M_i(v')|-i\hcbar\stackrel{\leftarrow}{D}_{\rho}\Gamma\hc |
M_j^{**}(v)\rangle =}
\nonumber \\
&&\sqrt{M_S M_{P_{3/2}}} \mbox{Tr} \lbrack
\hat M_i^{\dagger}\frac{1+\vslash\,'}{2}\Gamma\frac{1+\vslash}{2}
\hat M_j^{**\alpha} (f_1 g_{\alpha\rho} + f_2v'_{\alpha} v_{\rho}
+ f_3v'_{\alpha}v'_{\rho} + f_4v'_{\alpha}\gamma_{\rho})\rbrack \,.
\eea
Here $f_{1-4}$ are functions of $v\cdot v'$. One gets a first constraint
 on these
unknown form--factors by contracting with $v'^{\rho}$ and using the
equation of motion $\hcbar\stackrel{\leftarrow}{D}\cdot v'$ = 0:
\beq
   f_1 + f_2 v\cdot v' +f_3 - f_4 = 0.
\eeq
A second constraint is obtained by taking the derivative
\bea
\lefteqn{
\partial_{\rho}\langle M_i(v')|-i\hcbar\Gamma\hc |M_j^{**}(v)\rangle}
\nonumber\\
&&= (\bar\Lambda 'v'-\bar\Lambda v)_{\rho}\langle M_i(v')|\hcbar\Gamma
\hc|M_j^{**}(v)\rangle \nonumber \\
&&=\langle M_i(v')|-i\hcbar\stackrel{\leftarrow}{D}_{\rho}\Gamma\hc -
i\hcbar\Gamma\stackrel{\rightarrow}{D}_{\rho}\hc |M_j^{**}(v)\rangle\,.
\eea
Contracting with $v^{\rho}$, employing the equation of motion $v\cdot
D\hc$ = 0 and comparing with (16), a second constraint is obtained
\beq
   f_2 + f_3 v\cdot v' - f_4 = \lbrack\bar\Lambda'v\cdot v'-\bar
\Lambda\rbrack\xi_{3/2}^*(v\cdot v')\,.
\eeq
{}From (17) and (19) we get
\beq
   f_1(1)=-(\bar\Lambda'-\bar\Lambda)\xi_{3/2}^*(1)\,.
\eeq
The matrix element of the second term in (6) can be expressed with the
help of equ. (18) as
\bea
\lefteqn{
-i\langle M_i(v')|\hcbar(\stackrel{\leftarrow}{\Dslash}\gamma_{\mu}-
\gamma_{\mu}\stackrel{\rightarrow}{\Dslash})\hc |M_j^{**}(v)\rangle =}
 \\
&&-2i\langle M_i(v')|\hcbar\stackrel{\leftarrow}{D}_{\mu}\hc |
M_j^{**}(v)\rangle - (\bar\Lambda 'v'-\bar\Lambda v)_{\rho}
\langle M_i(v')|\hcbar\gamma_{\mu}\gamma^{\rho}\hc |M_j^{**}(v)\rangle
\,.\nonumber
\eea
The first term on the r.h.s. is given by a formula like (16) with
$\Gamma=1$, where it
can be seen that for $v_{\mu}=v'_{\mu}$ only the term with $f_1$
survives because of the orthogonality relation $\hat M_j^{**\alpha}
v_{\alpha}$ = 0. But from (20) one sees that at $v\cdot v'=1$, $f_1$
can be expressed only in terms of $\xi_{3/2}^*(1)$. This is essentially the
explanation for our simple final result.

   Now for the contribution of the non--leading terms in the HQET
Lagrangian
(15). The $\hcbarv (iD)^2\hc$ term just "renormalizes" $\xi_{3/2}^*$ by
effectively changing
\beq
   \xi_{3/2}^*(v\cdot v') \to \xi_{3/2}^*(v\cdot v') + \frac{1}{m_c}
\eta(v\cdot v')
\eeq
and the chromomagnetic term $\hcbarv\sigma\cdot G\hc$ gives a contribution
to the matrix element of interest equal to
\bea
\sqrt{M_SM_{P_{3/2}}}\frac{1}{m_c}\mbox{Tr}\lbrack\hat M_i^{\dagger}
\frac{1+\vslash\,'}{2}\gamma_{\mu}\frac{1+\vslash}{2}\sigma^{\rho\lambda}
\frac{1+\vslash}{2}\hat M_j^{**\alpha}v'_{\alpha}(\chi_1\sigma_{\rho\lambda}
+i\chi_2(v'_{\rho}\gamma_{\lambda}-v'_{\lambda}\gamma_{\rho}))\rbrack
\nonumber \\
+\sqrt{M_SM_{P_{3/2}}}\frac{1}{m_c}\mbox{Tr}\lbrack\hat M_i^{\dagger}
\frac{1+\vslash\,'}{2}\sigma^{\rho\lambda}\frac{1+\vslash\,'}{2}
\gamma_{\mu}\frac{1+\vslash}{2}\hat M_j^{**\alpha}v'_{\alpha}
(\chi_1\sigma_{\rho\lambda}-i\chi_2(v_{\rho}\gamma_{\lambda}-v_{\lambda}
\gamma_{\rho}))\rbrack\,.
\eea
$\eta,\chi_1$ and $\chi_2$ are real unknown functions. There are thus
five unknown form factors besides the Isgur--Wise function $\xi_{3/2}^*$
which determine the matrix elements of the (approximatively) conserved
vector current $\bar c\gamma_{\mu}c$ in the heavy quark effective
theory up to ${\cal O}(1/m_c)$\footnote{Similar results were obtained
also in the Ref.\cite{MR}.}.
In the expressions for these matrix elements we insert $v'_{\mu}$  as a
function of $v_{\mu}$ and $k_{\mu}$ from (7), the meson masses $M_S$
and $M_{P_{3/2}}$ are expressed in terms of $\bar\Lambda$ and
$\bar\Lambda'$
and the result is expanded in powers of $\frac{k}{m_c}$. In this expansion,
one power of $k_{\mu}$ counts as much as one power of $\bar\Lambda$
, $\bar\Lambda'$, $f_{1-4}$, $\eta$ or $\chi_{1,2}$ as they are of the
same order of magnitude. The leading terms of this expansion look as follows:
\bea
\lefteqn{\langle D^*(v',\epsilon_2)|\tilde J_{\mu}^{heavy}|
D_2^*(v,\epsilon_1)\rangle} \nonumber\\
&&=\sqrt{M_SM_{P_{3/2}}}\left( -\frac{2}{m_c}
\epsilon_1^{\alpha\lambda}
\epsilon^*_{2\alpha}k_{\lambda}v_{\mu} + 2\frac{\bar\Lambda-\bar\Lambda'}
{m_c}\epsilon_1^{\mu\alpha}\epsilon^*_{2\alpha}\right)\xi_{3/2}^*(1)\,,
\\
\lefteqn{\langle D^*(v',\epsilon_2)|\tilde J_{\mu}^{heavy}|D_1(v,\epsilon_1)
\rangle =\sqrt{M_SM_{P_{3/2}}} \frac{2}{\sqrt{6}m_c}i\epsilon_{\alpha
\beta\gamma\mu}\epsilon_1^{\alpha}
\epsilon^{*\beta}_2k^{\gamma}\xi_{3/2}^*(1)\,,}
\\
\lefteqn{\langle D(v')|\tilde J_{\mu}^{heavy}|D_2^*(v,\epsilon_1)
\rangle =\sqrt{M_SM_{P_{3/2}}} \frac{1}{m_c^2}i\epsilon_{\alpha
\beta\gamma\mu}\epsilon_1^{\beta\rho}k_{\rho}v^{\alpha}k^{\gamma}
\xi_{3/2}^*(1)\,,}
\\
\lefteqn{\langle D(v')|\tilde J_{\mu}^{heavy}|D_1(v,\epsilon_1)
\rangle\nonumber}\\
&&=\sqrt{M_SM_{P_{3/2}}}\left( \frac{4}{\sqrt{6}m_c}
(k\cdot\epsilon_1)v_{\mu} - \frac{4}{\sqrt{6}m_c}
(\bar\Lambda-\bar\Lambda')\epsilon_{1\mu}\right)\xi_{3/2}^*(1)\,.
\eea
   It is apparent that all the matrix elements (24-27) can be expressed
only in terms of
$\xi_{3/2}^*(1)$ and that none of the five subleading form factors
contributes. An important role is played here by equ. (10) which
prevents the appearance of terms proportional to the derivative of
$\xi_{3/2}^*(x)$ at $x=1$.
   These matrix elements exhibit approximate gauge invariance (up to
${\cal O}(1/m_c)$), as shown in (14). In practice this is slightly
inconvenient, so that we will henceforth replace
$\bar\Lambda-\bar\Lambda'$ by $v\cdot k$ in the above equations,
which only changes the result by a next-to-leading quantity, as is
apparent from (9).
After this change our matrix elements are explicitly gauge invariant.

\section*{3.Results}

   The four possible electromagnetic decays of the  $s_{\ell}
^{\pi_{\ell}} = \frac{3}{2}^+$ charmed mesons to the $s_{\ell}
^{\pi_{\ell}} = \frac{1}{2}^-$ charmed mesons have, in the infinite mass
limit,
the following multipole content: $D_2^*\stackrel{E1,\,M2}{\to}D^*$,
$D_2^*\stackrel{M2}{\to}D$, $D_1\stackrel{E1,\,M2}{\to}D^*$,
$D_1\stackrel{E1}{\to}D$.

   The $D_2^*\to D^*$ transition can in general proceed also through a
$E3$ mode, as far as angular momentum and parity are concerned.
In the infinite mass limit this is forbidden. However, this
prediction is not very likely to be easily tested experimentally.

   More interesting are the heavy quark symmetry predictions for the
ratios of partial amplitudes for these decays:
\bea
 A_{E1}(D_2^*\to D^*\gamma)\,&:&\,A_{E1}(D_2^*\to D\gamma)\,:\,
 A_{E1}(D_1\to D^*\gamma)\,:\,A_{E1}(D_1\to D\gamma)\,\,
\nonumber\\
&=&\sqrt{3}\,:\,0\,:\,1\,:\,\sqrt{2}   \\
 A_{M2}(D_2^*\to D^*\gamma)\,&:&\,A_{M2}(D_2^*\to D\gamma)\,:\,
 A_{M2}(D_1\to D^*\gamma)\,:\,A_{M2}(D_1\to D\gamma)\,\,
\nonumber\\
&=&\sqrt{3}\,:\,\sqrt{2}\,:\,\sqrt{5}\,:\,0\,,
\eea
which can e.g. be obtained along the lines of Ref.\cite{IW}.
Using (5) and the matrix elements (24-27) we obtain the following
one-photon widths, which automatically satisfy the above relations:
\bea
 \Gamma(D_2^*\to D^*\gamma) &=& \frac{4\alpha}{3}\frac{M_{D^*}}{M_{D_2^*}}
|\vec k|^3 (e_qF+e_Q\frac{\xi_{3/2}^*(1)}{m_c})^2 + \frac{3\alpha}{5}
\frac{M_{D^*}}{M_{D_2^*}}|\vec k|^5(e_qG)^2\,,
\\
 \Gamma(D_2^*\to D\gamma) &=& \frac{2\alpha}{5}
\frac{M_D}{M_{D_2^*}}|\vec k|^5(e_qG+e_Q\frac{\xi_{3/2}^*(1)}{2m_c^2})^2\,,
\\
 \Gamma(D_1\to D^*\gamma) &=& \frac{4\alpha}{9}\frac{M_{D^*}}{M_{D_1}}
|\vec k|^3 (e_qF+e_Q\frac{\xi_{3/2}^*(1)}{m_c})^2 + \alpha
\frac{M_{D^*}}{M_{D_1}}|\vec k|^5(e_qG)^2\,,
\\
 \Gamma(D_1\to D\gamma) &=& \frac{8\alpha}{9}\frac{M_D}{M_{D_1}}
|\vec k|^3 (e_qF+e_Q\frac{\xi_{3/2}^*(1)}{m_c})^2\,.
\eea
   Here $e_q$ and $e_Q \,(= \frac{2}{3}$ for Q = c) are the light and
the heavy quark electric charges and $F$ , $G$ are given by
\bea
   F&=&f_1-f_2+\frac{v\cdot k}{2\Lambda_{\chi}}(g_1+2g_2+4g_3)\,, \\
   G&=&\frac{1}{\Lambda_{\chi}}(g_1+2g_2)\,.
\eea
The heavy quark contribution to the $M2$ partial width has been written
only for the transition $D_2^*\to D\gamma$, since it is only in this case
that it can be expressed solely in terms of $\xi_{3/2}^*(1)$. For all
the other decays, it depends also on some of the five unknown subleading
form factors.

   We will use in the following the ISGW value\cite{5}
\beq
     \xi_{3/2}^*(1)=0.584\,.
\eeq
 This corresponds to a transition $c\to c$ and is slightly different
from the corresponding value for a $b\to c$ transition, of 0.537
\footnote{The relation to the function $\tilde\tau_{3/2}$ used in \cite{5}
is $\xi_{3/2}^*=\sqrt{3}\tilde \tau_{3/2}$.} \footnote{Note that what we
call $\xi_{3/2}^*$ is not the usual Isgur--Wise function, but rather its
renormalization group invariant version, which includes a scaling factor
with the respective velocity dependent anomalous dimension and has a
logarithmic dependence on the heavy quark mass. This in part
explains the two different numerical values quoted above.}. A similar value
is also obtained from a recent QCD sum rule calculation\cite{PAVER}.
 Although this value is the result of a model calculation, $\xi_{3/2}^*(1)$
can be obtained
experimentally, from a study of the differential cross-section for
the semileptonic decays of $B$ mesons into p--wave D mesons. In this sense
the relations to be derived below are truly model--independent.

   One first set of predictions is nothing else but the mass corrections
to the ratios (28-29). In the approximation that $F$ and $G$ are constants
independent of the photon energy, these are:
\bea
 \Gamma_{E1}(D_{s2}^{*+}\to D_s^{*+}\gamma)\,&:&\,\Gamma_{E1}(D_{s2}^{*+}
\to D_s^+\gamma)\,:\,\Gamma_{E1}(D_{s1}^+\to D_s^{*+}\gamma)\,:\,
\Gamma_{E1}(D_{s1}^+\to D_s^+\gamma)\,\,
\nonumber\\
&=&3.694\,:\,0\,:\,1\,:\,4.071   \\
 \Gamma_{M2}(D_{s2}^{*+}\to D_s^{*+}\gamma)\,&:&\,\Gamma_{M2}(D_{s2}^{*+}
\to D_s^+\gamma)\,:\,\Gamma_{M2}(D_{s1}^+\to D_s^{*+}\gamma)\,:\,
\Gamma_{M2}(D_{s1}^+\to D_s^+\gamma)\,\,
\nonumber\\
&=&0.965\,:\,2\,:\,1.212\,:\,0\,.
\eea
and similarly for the $D^0$ mesons:
\bea
 \Gamma_{E1}(D_2^{*0}\to D^{*0}\gamma)\,&:&\,\Gamma_{E1}(D_2^{*0}
\to D^0\gamma)\,:\,\Gamma_{E1}(D_1^0\to D^{*0}\gamma)\,:\,\Gamma_{E1}
(D_1^0\to D^0\gamma)\,\,
\nonumber\\
&=&3.539\,:\,0\,:\,1\,:\,4.022   \\
 \Gamma_{M2}(D_2^{*0}\to D^{*0}\gamma)\,&:&\,\Gamma_{M2}(D_2^{*0}
\to D^0\gamma)\,:\,\Gamma_{M2}(D_1^0\to D^{*0}\gamma)\,:\,\Gamma_{M2}
(D_1^0\to D^0\gamma)\,\,
\nonumber\\
&=&0.964\,:\,2\,:\,1.125\,:\,0\,.
\eea

   In the corresponding one pion decay case a small mixing between the
$D_1$ and the
axial vector member of the  $s_{\ell}^{\pi_{\ell}} = \frac{1}{2}^+$
multiplet could significantly alter the predicted amplitude ratio.
However, in our case this is no longer true. Both these states can decay
to the s--wave mesons through a $E1$ mode, and as far as the $M2$ channel is
concerned, it is only the decay of the vector member of the
$s_{\ell}^{\pi_{\ell}}=\frac{1}{2}^+$ multiplet which is forbidden by
large mass limit selection rules.

   It is clear that without any knowledge of the constants $F$ and $G$
no further advance can be made. Since the value (36) has been obtained
by using the constituent quark model of Isgur, Scora, Grinstein and Wise,
consistency requires that we use the same model for determining them.
Although the authors of Ref.\cite{6} made use of their model only to
calculate matrix elements
of heavy quark current bilinears, it can give matrix elements
of light quark currents as well, thereby providing a way for evaluating our
parameters $F$ and $G$. The model is well tested and is known to give
reliable results for both the s-- and p--wave charmed mesons. Furthermore,
it provides a simple way for including $SU(3)$ violating effects induced
by a constituent strange quark mass $m_s$ different from $m_u$, $m_d$.

   To determine $F$ and $G$ for the $D^0$ meson case, consider the
following matrix element of the light quark current $\bar u\gamma_{\mu} u$,
which according to the effective lagrangian (5) is given by
\beq
 \langle D^0(v)|\bar u\gamma_{\mu} u|D_1^0(v,\epsilon)\rangle =
4\sqrt{M_{D^0}M_{D_1^0}}\lbrack -(v\cdot k)\epsilon_{\mu} +
(\epsilon\cdot k)v_{\mu}\rbrack F_{D^0}\,.
\eeq
The same matrix element is written as
\beq
 \langle D^0(v)|\bar u\gamma_{\mu} u|D_1^0(v,\epsilon)\rangle =
\sqrt{M_{D^0}M_{D_1^0}}\lbrack -2.673 (v\cdot k)\epsilon_{\mu} + 2.396
(\epsilon\cdot k)v_{\mu}\rbrack \,\,(GeV)\,,
\eeq
where we make use of the relations given in the Appendix B of Ref.\cite{6}.
Here some care is needed, as the light quark in a $D$ meson is an
antiquark. Therefore the direct application of the relations of Ref.
\cite{6} gives rather the matrix element between the corresponding
$\bar D$ mesons. However, these can be related by crossing to the matrix
elements of interest.
Unfortunately, the transition matrix element (42) is not gauge invariant.
This is a typical problem of
constituent quark model calculations of radiative transition matrix
elements (see e.g. \cite{FKR}) and we will deal with it by adopting as the
value of the coupling $4F$ the average of the two numbers in (42), with
an error given by their difference. This should give at the same time some
measure of the model dependence of the result. Thus we take
\beq
       F_{D^0} = (0.634\pm 0.034)\,\,GeV^{-1}
\eeq
In obtaining this number we have used a value $\kappa=1$ for the
"relativistic correction factor" $\kappa$ (see Ref.23 in \cite{5}).
Adopting $\kappa = 0.7$ lowers $F$ by a factor of about 0.5.
   Similarly the value of $G$ can be obtained by calculating the matrix
element
\beq
 \langle D^0(v)|\bar u\gamma_{\mu}u|D_2^{*0}(v,\epsilon)\rangle =
-2i\sqrt{M_{D^0}M_{D_2^{*0}}}G_{D^0}\epsilon_{\mu\nu\lambda\rho}
\epsilon^{\nu\alpha}k_{\alpha}v^{\lambda}k^{\rho}
\eeq
with the result
\beq
  G_{D^0}= - 2.168\,\,GeV^{-2}\,.
\eeq
Inserting these values
for $F_{D^0}$ and $G_{D^0}$ in the rate formulae (30-33), we obtain
the partial widths in Table 1.  Here a charmed quark mass
$m_c=1.82\,\,GeV$ has been used.

   A similar calculation gives the following values for the corresponding
couplings for the $D_s^+$ mesons:
\bea
  F_{D_s^+}&=&(0.493\pm 0.083)\,\,GeV^{-1}\\
  G_{D_s^+}&=&-1.254\,\,GeV^{-2}\,.
\eea
In obtaining these numbers the following values for the oscillator
strength parameters of the model have been used: $\beta_S=0.40\,\,GeV$
and $\beta_P=0.35\,\,GeV$. For the flavour content $s\bar c$
these parameters are not available in the Ref.\cite{6} and therefore
we simply took them slightly larger than for $u\bar c$, as seems to
be the case for the light mesons. The mass of the tensor state
$D_{2s}^{*+}$ has been taken equal to $M_{D_{2s}^{*+}}=2564\,MeV$
\cite{ISGUR}. The partial rates for this case are
also shown in the Table 1. Several remarks are in order about these
results:
\begin{itemize}

\item The heavy quark contribution amounts to about 34\% in absolute
value in the $E1$ amplitude for the $D^0$ mesons' case and 56\% for the
$D_s^+$ case. In the former it has an enhancement effect while in the
latter case it contributes with an opposite sign. On the other hand,
in the $M2$ amplitude for the $D_2^*\to D$ transition the heavy quark
current contributes negligibly (under 4\%). Very likely, the same is true
for all the other $M2$ amplitudes.

\item Our estimates for the total one-photon widths agree well with
the predictions in \cite{SIMILAR2} for the decay
$D_{2s}^{*+} \to D_s^{*+}$ and are lower for the same decay of the
$D^0$ meson. Compared to the \cite{SIMILAR1}, our estimates for
$\Gamma(D_2^*\to D\gamma)$ are down by a factor of 1.8 and 5.5 for
the $D^0$ and $D_s^+$ cases respectively. Also, we get a value
for $\Gamma(D_2^{*0}\to D^{*0}\gamma)$ which is lower by a factor of
5.2 than the one given in \cite{SIMILAR3}. As a general
common point we mention the strong suppression of the decay
widths of the $D_s^+$ mesons compared to the $D^0$ mesons.

\item These estimates yield branching ratios of about (0.9--1.5)\%
for the radiative decays of the $D^0$ mesons and larger than
(0.02--0.4)\% for the $D_{s1}^+$, for which an upper limit for the total
width of 3.9 MeV exists\cite{ISGUR}.

\end{itemize}

   The largest source of errors in our approach seems to be the
neglect of higher order terms in the effective lagrangian (5), with more
derivatives on the photon field. Such terms would be a source of
additional photon-energy dependence of the couplings $F$ and $G$.
An estimate for this dependence can be obtained by formally setting
$g_3=0$ in (34). Then the dependence of $F$ with the photon energy
can be expressed simply in terms of $G$. This yields a variation of $F$
among the three allowed $E1$ transitions of about 15 \%, with the
attendant corrections to the amplitude ratios (37) and (39).

   The finite mass corrections are only significant in the $E1$ amplitudes,
and we have estimated them to contribute up to 10\% of the amplitude.
Much more likely are important the corrections due to $SU(3)$ violation.
These can be calculated in a systematical way\cite{1} using the heavy
hadron chiral perturbation theory. We have only evaluated them in an
effective way, from a constituent quark model, to be of about 25\% for
$F$ and 50\% for $G$.

   Similar methods can be applied to the description of the
electromagnetic decays of the members of the $s_{\ell}^{\pi_{\ell}}
 = \frac{1}{2}^+$ multiplet of charmed mesons and of the similar
excited B mesons. Since the former are expected to be quite broad
($\Gamma \sim 200\, MeV$) because of the pionic decay mode, their
photonic branching ratia are surely very small. As for the latter, in
the view of the absence of any experimental data about these states,
such applications must be reserved for the future.

\vfill\eject
$$ \vbox{\offinterlineskip
\def\tablerule{\noalign{\hrule}}
\hrule
\halign {\vrule#& \strut#&
\ \hfil#\hfil& \vrule#&
\ \hfil#\hfil& \vrule#&
\ \hfil#\hfil& \vrule#&
\ \hfil#\hfil& \vrule#&
\ \hfil#\hfil\ & \vrule# \cr
\tablerule
height10pt && \omit && \omit && \omit && \omit && \omit &\cr
&& \quad Decay \quad && \quad $\Gamma_{E1}\,\,(keV)$ \quad
&& \quad $\Gamma_{M2}\,\,(keV)$ \quad && \quad $\Gamma_{tot}\,\,(keV)$
\quad && \quad $|\vec k|\,\,(MeV)$ \quad & \cr
height10pt && \omit && \omit && \omit && \omit && \omit &\cr
\tablerule
height10pt && \omit && \omit && \omit && \omit && \omit &\cr
&& $\Gamma(D_1^0 \to D^0 \gamma)$ && $(245 \pm 18) $ &&
   $ 0 $ && $ (245 \pm 18) $ && $ 494.93 $ &\cr
height10pt && \omit && \omit && \omit && \omit && \omit &\cr
&& $\Gamma(D_1^0 \to D^{*0} \gamma)$ && $( 60 \pm 5) $ &&
   $ 101 $ && $ (161 \pm 5) $ && $ 381.05 $ &\cr
height10pt && \omit && \omit && \omit && \omit && \omit &\cr
&& $\Gamma(D_2^{*0} \to D^{*0} \gamma)$ && $( 222 \pm 16) $ &&
   $ 87 $ && $(309 \pm 16) $ && $ 410.38 $ &\cr
height10pt && \omit && \omit && \omit && \omit && \omit &\cr
&& $\Gamma(D_2^{*0} \to D^0 \gamma)$ && $ 0  $ &&
   $ 181 $ && $ 181 $ && $ 522.63 $ &\cr
height10pt && \omit && \omit && \omit && \omit && \omit &\cr
\tablerule
height10pt && \omit && \omit && \omit && \omit && \omit &\cr
&& $\Gamma(D_{s1}^+ \to D_s^+ \gamma)$ && $(1.6 \pm 2.3)$ &&
   $ 0 $ && $ (1.6 \pm 2.3) $ && $ 503.77 $ &\cr
height10pt && \omit && \omit && \omit && \omit && \omit &\cr
&& $\Gamma(D_{s1}^+ \to D_s^{*+} \gamma)$ && $(0.4\pm 1.0)$ &&
   $ 10.0 $ && $ (10.4 \pm 1.0) $ && $ 389.97 $ &\cr
height10pt && \omit && \omit && \omit && \omit && \omit &\cr
&& $\Gamma(D_{s2}^{*+} \to D_s^{*+} \gamma)$ && $( 1.4 \pm 2.0)$ &&
   $ 8.0 $ && $ ( 9.4 \pm 2.0) $ && $ 413.56 $ &\cr
height10pt && \omit && \omit && \omit && \omit && \omit &\cr
&& $\Gamma(D_{s2}^{*+} \to D_s^+ \gamma)$ && $ 0 $ &&
   $ 16.0 $ && $ 16.0 $ && $ 526.12 $ &\cr
height10pt && \omit && \omit && \omit && \omit && \omit &\cr
\tablerule}} $$
{\bf Table 1.}The calculated partial widths for the radiative
decays of the
$s_{\ell}^{\pi_{\ell}} = \frac{3}{2}^+$ p--wave charmed
mesons to s-wave charmed mesons.
In the last column the corresponding
photon momentum is given.

\end{document}